\documentclass[letterpaper]{article}
\usepackage[draft]{aaai25} 
\usepackage{times} 
\usepackage{helvet} 
\usepackage{courier} 
\usepackage[hyphens]{url} 
\usepackage{graphicx} 
\usepackage{graphicx}  
\usepackage{natbib}  
\usepackage{caption}  
\usepackage{multirow}
\usepackage{xcolor} 
\usepackage{amsmath} 
\usepackage{amssymb}  

\usepackage{todonotes}
\presetkeys{todonotes}{inline}{}

\usepackage{listings}
\usepackage{csvsimple}

\lstdefinelanguage{JavaScript}{
  keywords={typeof, new, true, false, catch, function, return, null, catch, switch, var, if, in, while, do, else, case, break},
  keywordstyle=\color{blue}\bfseries,
  ndkeywords={class, export, boolean, throw, implements, import, this},
  ndkeywordstyle=\color{darkgray}\bfseries,
  identifierstyle=\color{black},
  sensitive=false,
  comment=[l]{//},
  morecomment=[s]{/*}{*/},
  commentstyle=\color{green!50!black}\ttfamily,
  stringstyle=\color{red}\ttfamily,
  morestring=[b]',
  morestring=[b]"
}

\title{
Benchmarking LLM Code Generation for Audio Programming with Visual Dataflow Languages}

\author{
       William Zhang\textsuperscript{\rm 1},
       Maria Leon\textsuperscript{\rm 2},\\
       Ryan Xu, Adrian Cardenas, Amelia Wissink, Hanna Martin, Maya Srikanth, Kaya Dorogi, Christian Valadez, Pedro Perez, Citlalli Grijalva\textsuperscript{\rm 1},\\
       Corey Zhang\textsuperscript{\rm 3},
       Mark Santolucito\textsuperscript{\rm 2}
    }
\affiliations {
       \textsuperscript{\rm 1}Columbia University, New York, USA\\
       \textsuperscript{\rm 2}Barnard College, Columbia University, New York, USA\\
       \textsuperscript{\rm 3}Eastlake High School, Sammamish, USA\\
       
       zhang.william@columbia.edu,
       msl2218@barnard.edu,
       \{rx2189, ac4982, afw2122, hem21620, ms6198, kad2220, cdv2119, pap2153, cg3236\}@columbia.edu,
       s-cozhang@lwsd.org,
       msantolu@barnard.edu
}

\begin{document}

\maketitle

\begin{abstract}
Node-based programming languages are increasingly popular in media arts coding domains.
These languages are designed to be accessible to users with limited coding experience, allowing them to achieve creative output without an extensive programming background.
Using LLM-based code generation to further lower the barrier to creative output is an exciting opportunity.
However, the best strategy for code generation for visual node-based programming languages is still an open question.
In particular, such languages have multiple levels of representation in text, each of which may be used for code generation.
In this work, we explore the performance of LLM code generation in audio programming tasks in visual programming languages at multiple levels of representation. 
We explore code generation through metaprogramming code representations for these languages (i.e., coding the language using a different high-level text-based programming language), as well as through direct node generation with JSON.
We evaluate code generated in this way for two visual languages for audio programming on a benchmark set of coding problems.
We measure both correctness and complexity of the generated code.
We find that metaprogramming results in more semantically correct generated code, given that the code is well-formed (i.e., is syntactically correct and runs). 
We also find that prompting for richer metaprogramming using randomness and loops led to more complex code.
\\
\\ \textbf{KEYWORDS}
\\ large language models, audio programming, node based languages, code generation
    
\end{abstract}

\section{Introduction}
Large Language Models (LLMs) have shown promising results for code generation. 
To date, significant research has focused on their application to traditional software engineering-style coding problems~\cite{codex,isCodeCorrect}.
These applications of LLM code generation have been focused on software with impressive levels of complexity that can be used in real-world, professional software systems deployments.
However, one of the most exciting parts of LLM code generation has been the adoption by non-professional programmers into their workflow.

In support of a diversity of users of LLM code generation, we explore the use of LLMs for coding in the context of visual dataflow programming languages for audio. 
Visual dataflow languages for the arts, such as MaxMSP for audio~\cite{maxmsp}, 
Grasshopper for 3D modelling~\cite{grasshopper}, and shader graphs for video game visual artists~\cite{shadergraphs}, have allowed many non-traditional programmers to leverage the power of computation in their work.
Just as LLM code generation is delivering enormous benefit to traditional software engineers~\cite{jimenez2023swe}, it is critical that we also enable LLM code generation to benefit non-traditional programmers in their preferred programming environments.

If LLM code generation focuses only on text-based languages, which may lie outside the area of expertise of some users, the generated code then functions effectively as a blackbox for some users—restricting the user who prefers visual programming. 
Code comprehensibility and the ability to modify the generated computational artifacts is critical in the world of automatic program construction~\cite{santolucito2018programming, santolucito2018pbe,santolucito2019live,hempel2019sketch,mcnuttCHI2023,ferdowsi2023coldeco}.
Generating code in languages that allow for users to make manual edits can help preserve the flexibility and expressiveness of their workflow.

The main contributions of this work are as follows:

\begin{enumerate}
    \item Propose a benchmark set for audio digital signal processing (DSP) and use it to evaluate LLM code generation.
    \item Evaluate 600 code generations over two languages—one of our own design and one industry standard language for audio DSP—for visual node-based audio programming across three different levels of code representation on this benchmark set.
    \item Define a metric that measures if LLM-generated code is semantically correct, given that it is well-formed.
    \item Provide an analysis of LLM code generation comparing direct JSON generation to metaprogramming for visual dataflow languages that may guide future work in the area, finding that metaprogramming results in more semantically correct code given it is well-formed.

\end{enumerate}

\section{Background and Related Work}

The topic of using LLM for creative work has seen an explosion of interest.
There have been new languages proposed that use LLMs as primitives in the language. For example, Jigsaw is a visual language for the construction of LLM pipelines~\cite{lin2023jigsaw}.
Other work has focused on the intersection of media arts code and LLM code generation, for example in P5.js~\cite{wang2024exploring}, a text-based language for web-based animations.
Spellburst is an interface for creative coding assisted by LLMs~\cite{spellburst}, which gives users a visual interface for tracking their interaction with the LLM and moving between the LLM modality, and the code modality of creation.

In the domain of audio programming with visual languages, designing an LLM-based assistant for MaxMSP is a crucial step to make the language more accessible to users with limited programming experience \cite{zhang2024harmonyassist}.
However, there is limited work in the best methods of LLM code generation specifically for visual languages, especially in the creative coding space. 
An important part of LLM-based code generation is that the generated code be understandable~\cite{ferdowsi2023coldeco}. Generating quality code for visual languages is an important direction, as some users find the visual language to be a more understandable code format than text-based languages.

We focus on visual dataflow languages for audio programming. 
We explore two languages in this context.

First, we use MaxMSP~\cite{maxmsp}, a visual dataflow language designed for interactive audiovisual projects.
MaxMSP is one of the most popular languages in the computer-based music community.
To complement the visual language of MaxMSP, we also generate code in MaxPy, a Python library for \textbf{metaprogramming} (writing code that generates code) in MaxMSP.

Second, we use the Web Audio API, a JavaScript API and one of the most popular frameworks for in-browser audio programming.
To complement the text-based Web Audio API, we also generate code in the visual programming language Wavir, a node-based editor for Web Audio. 

We provide here a short introduction to each language, highlighting the language features that are most relevant to our investigation of code generation. 

\begin{figure}[t!]
    \centering
    \includegraphics[width=.99\columnwidth]{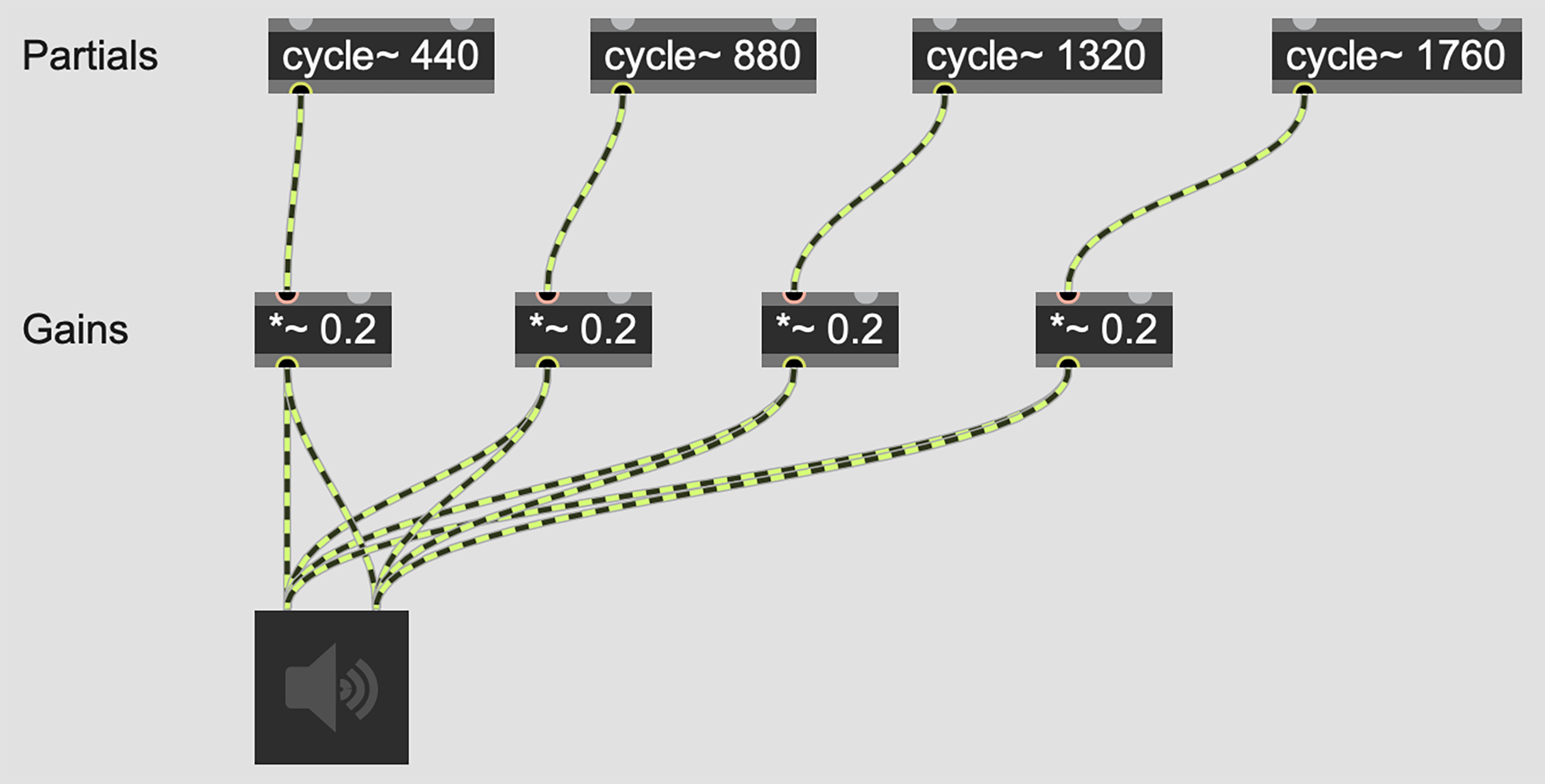}
    \caption{A screenshot of a MaxMSP patch for additive synthesis.}
    \label{fig:maxmsp}
\end{figure}

\begin{figure*}[t!]
    \centering
    \includegraphics[width=\textwidth]{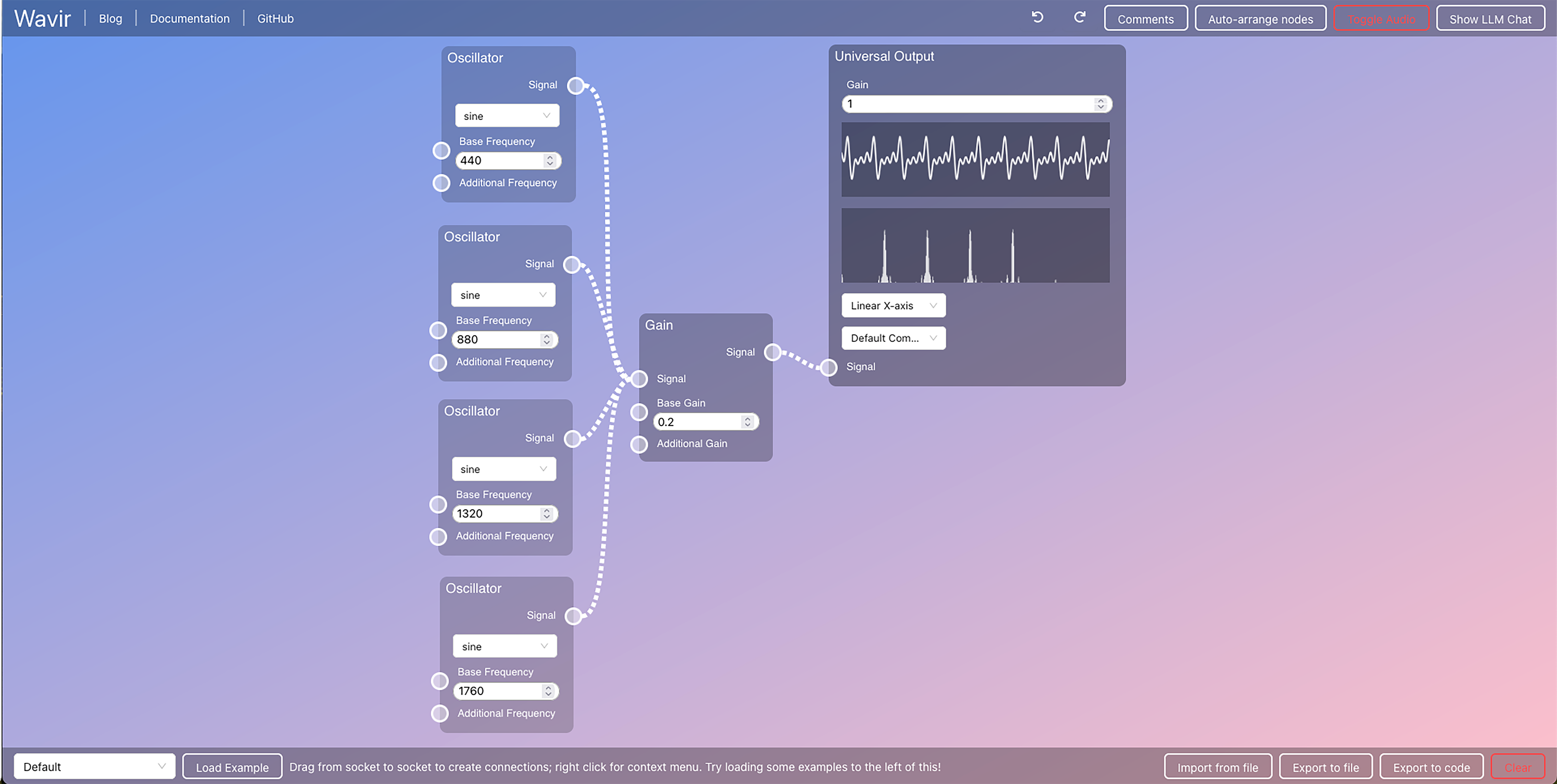}
    \caption{A screenshot of the Wavir visual language for Additive synthesis.}
    \label{fig:wavvr}
\end{figure*}

We also give example code snippets for the same program (additive synthesis on a sine wave at 440 Hz with 3 partials above the fundamental) in each language as shown in Fig.~\ref{fig:maxmsp}, Fig.~\ref{fig:python-code}, Fig.~\ref{fig:python-rich-code}, Fig.~\ref{fig:wavvr}, Fig.~\ref{fig:web-audio-code}, and Fig.~\ref{fig:web-audio-rich-code}.

\textbf{MaxMSP} is a visual programming language, with each project represented in an interface of nodes and connections called a MaxMSP program (called a ``patch" in the MaxMSP community).
This visual nature makes the language accessible to users, but this visual-first representation also presents challenges for existing text-based LLM code generation. MaxMSP files are saved in JSON format that captures objects, connections, and spatial layout (the placement of nodes). 

\textbf{MaxPy}~\cite{maxpy} is a Python package that allows users to create Max patches in Python through metaprogramming. Users can use Python to programmatically construct MaxMSP patches—a MaxPy program generates a JSON file that is a MaxMSP patch.

\textbf{Wavir} is a visual dataflow language of our own design that complements Web Audio and is inspired by MaxMSP~\cite{wavir}. It is available at \url{wavir.io}.  Wavir's semantics are approximately the same as MaxMSP. Nodes can be sources of sound (oscillators and noise), processors (filters or gains), or constants to modify other nodes, both as float numbers or as musical note names. There are also input nodes for keyboard inputs and output nodes that display frequency and time visualizers or feed the signal to Web Audio's universal output. Nodes are connected to each other and finally to the output to generate sound.
One key difference between Wavir and MaxMSP is that spatial layout is absent from Wavir's JSON encoding.

The \textbf{Web Audio API} used in Wavir is a JavaScript API for processing and synthesizing audio in web applications. This API provides a way to manipulate audio through a modular routing system consisting of audio nodes that when connected form an audio processing graph. Different nodes represent audio processing operations. The dataflow model of the Web Audio API can be represented visually as a directed graph, where nodes are vertices, and connections between nodes are edges. The Web Audio API functions similarly to MaxPy as a metaprogramming language for programmatically constructing audio DSP graphs in Wavir.

For both MaxPy and Web Audio, we also define and use a level of code generation called \textbf{rich code}, where we prompt for code with more complexity.
We define rich code as a level of metaprogramming that produces more sophisticated programs by taking advantage of features only available in the text-based representation. In our experiment, we prompt the LLM to use for loops and random functions in our rich code trials. While MaxMSP has a random number object, a metalanguage such as MaxPy can use Python's random library to generate random values directly instead of adding another node to the Max patch. Moreover, the ability to generate large amounts of repetitive code in MaxMSP with a Python loop is one of the motivating factors of behind MaxPy's creation~\cite{maxpy}.
For large Max patches, this is more efficient than placing each repeated node individually through the visual editor.

Examples of rich code compared with ``normal code" (i.e., text-based code that does not leverage metalanguage features) are shown in Fig.~\ref{fig:python-rich-code} vs. Fig.~\ref{fig:python-code} and Fig.~\ref{fig:web-audio-rich-code} vs. Fig~\ref{fig:web-audio-code}.

\begin{figure}
    \begin{lstlisting}[language=Python]
# MaxPy example
import maxpy as mp

p = mp.MaxPatch()

fundamental = p.place("cycle~ 440", num_objs=1, starting_pos=[0, 100])[0]
partial1 = p.place("cycle~ 880", num_objs=1, starting_pos=[0, 150])[0]
partial2 = p.place("cycle~ 1320", num_objs=1, starting_pos=[100, 150])[0]
partial3 = p.place("cycle~ 1760", num_objs=1, starting_pos=[200, 150])[0]

mix = p.place("*~ 0.2", num_objs=1, starting_pos=[250, 300])[0]
ez = p.place("ezdac~", num_objs=1, starting_pos=[150, 400])[0]

p.connect([fundamental.outs[0], mix.ins[0]])
p.connect([partial1.outs[0], mix.ins[0]])
p.connect([partial2.outs[0], mix.ins[0]])
p.connect([partial3.outs[0], mix.ins[0]])

p.connect([mix.outs[0], ez.ins[1]])
p.connect([mix.outs[0], ez.ins[0]])

p.save("additive.maxpat")
    \end{lstlisting}
    \caption{Normal MaxPy code for Additive synthesis}
    \label{fig:python-code}
\end{figure}

\begin{figure}
    \begin{lstlisting}[language=Python]
# Rich MaxPy example
import maxpy as mp
import random

p = mp.MaxPatch()

fundamental = 440
f = p.place(f"cycle~ {fundamental}", num_objs=1, starting_pos=[100, 100])[0]

num_partials = 3
partials = []
for i in range(1, num_partials + 1):
    partial_freq = fundamental * (i+1) + random.random(
        ) * 15 if i % 2 == 0 else fundamental * (i+1) - random.random() * 15
    partial = p.place(f"cycle~ {partial_freq}", num_objs=1, starting_pos=[100 + i*50, 100 + i*50])[0]
    partials.append(partial)

mix = p.place("*~ 0.2", num_objs=1, starting_pos=[300, 300])[0]
ez = p.place("ezdac~", num_objs=1, starting_pos=[400, 300])[0]

p.connect([f.outs[0], mix.ins[0]])
for partial in partials:
    p.connect([partial.outs[0], mix.ins[0]])

p.connect([mix.outs[0], ez.ins[0]])
p.connect([mix.outs[0], ez.ins[1]])

p.save("additive.maxpat")
    \end{lstlisting}
    \caption{Rich MaxPy code for Additive synthesis}
    \label{fig:python-rich-code}
\end{figure}

\begin{figure}
    \begin{lstlisting}[language=JavaScript]
// Web Audio API code example
const audioCtx = new (window.AudioContext || window.webkitAudioContext)();

const osc1 = audioCtx.createOscillator();
const osc2 = audioCtx.createOscillator();
const osc3 = audioCtx.createOscillator();
const osc4 = audioCtx.createOscillator();

osc1.frequency.setValueAtTime(440, audioCtx.currentTime); 
osc2.frequency.setValueAtTime(880, audioCtx.currentTime); 
osc3.frequency.setValueAtTime(1320, audioCtx.currentTime);
osc4.frequency.setValueAtTime(1760, audioCtx.currentTime);

const globalGain = audioCtx.createGain();
globalGain.gain.value = 0.0001;

osc1.connect(globalGain);
osc2.connect(globalGain);
osc3.connect(globalGain);
osc4.connect(globalGain);

globalGain.connect(audioCtx.destination);

osc1.start();
osc2.start();
osc3.start();
osc4.start();

globalGain.gain.setValueAtTime(0.25, audioCtx.currentTime); 
globalGain.gain.setTargetAtTime(0.0001, audioCtx.currentTime + 0.2, 1);
    \end{lstlisting}
    \caption{Web Audio API code example for Additive synthesis.}
    \label{fig:web-audio-code}
\end{figure}

\begin{figure}
    \begin{lstlisting}[language=JavaScript]
// Web Audio API rich code example
const audioCtx = new AudioContext();

const fundamental = 440;
const nPartials = 3;

const globalGain = audioCtx.createGain();
globalGain.gain.value = 0.0001;

let oscs = [];
for (let i = 0; i <= nPartials; i++) {
    const osc = audioCtx.createOscillator();
    
    if (i === 0) {
        osc.frequency.value = fundamental;
    } else {
        osc.frequency.value = (i % 2 === 0) ? 
            fundamental * (i + 1) - Math.random() * 15 : 
            fundamental * (i + 1) + Math.random() * 15;
    }

    oscs.push(osc);
    osc.connect(globalGain);
    osc.start();
}

globalGain.connect(audioCtx.destination);
globalGain.gain.setTargetAtTime(0.2, audioCtx.currentTime, 0.05);
globalGain.gain.setTargetAtTime(0.0001, audioCtx.currentTime + 0.2, 1);
\end{lstlisting}
    \caption{Web Audio API rich code example for Additive synthesis.}
    \label{fig:web-audio-rich-code}
\end{figure}

\section{Methodology}

We investigate the following research questions:

\textbf{RQ1}: How does the correctness (as measured by pass@k) of the generated code differ between metaprogramming, rich metaprogramming, and JSON as the target of LLM generation for visual dataflow programming languages for audio

\textbf{RQ2} How does the complexity of the generated code differ between metaprogramming, rich metaprogramming, and JSON as the target of LLM generation for visual dataflow programming languages for audio?

To investigate these questions, we divided our two audio programming languages of MaxMSP and Web Audio into three code generation methods each: JSON generation, metaprogramming, and rich metaprogramming. 

Thus, we experimented over six language categories:

\noindent \textbf{Max MSP}

1) MaxMSP JSON; 2) MaxPy; 3) MaxPy Rich Code

\noindent\textbf{Web Audio}

4) Wavir JSON; 5) Web Audio; 6) Web Audio Rich Code

We prompted LLM assistants to generate code using each of these categories for a defined benchmark set of sound projects and evaluated the output on correctness and complexity. We describe our assistants, benchmark set, prompting methods, and evaluation metrics in the sections below.

\subsection{Assistants Used}

For each language representation (MaxMSP JSON, MaxPy, Wavir JSON, Web Audio), we created a custom LLM assistant using the OpenAI Assistants API \cite{assistantsAPI}. These assistants used the gpt-4-0125-preview model. The purpose of using the Assistants API is to provide the LLM with familiarity in these languages, which it may not have out of the box. Our assistants were each provided with a knowledge base consisting of several code examples in their target generation language and a ``knowledge document". 

A \textbf{knowledge document} is an instruction file (e.g. txt, pdf) uploaded to an OpenAI Assitant's knowledge base through the Assistants API. It contains instructions specific to the Assistant's task, ensuring the assistant follows the outlined guidelines for formatting code or JSON, such that it will not generate any unsupported methods, nodes, or data values. The Assistant retrieves it on each run. Generally, a knowledge document is simply a prefix to every prompt. An example instruction from the MaxPy knowledge document is as follows: ``These are the ONLY methods of a MaxPatch object: patch = mp.MaxPatch(), patch.place, patch.connect, patch.save. Any other method will not compile."

\subsection{Benchmark Set}

We chose a set of 10 audio programming projects to test code generation across each language category. Our set includes 5 specific benchmarks and 5 creative benchmarks. Specific benchmarks test the implementation of highly-defined, low-level tasks, while creative benchmarks test open-ended creative code possibilities. 
Our benchmark set asks for implementations of the following sounds:

\noindent\textbf{Specific Benchmarks}
1) additive synthesis; 2) AM synthesis; 3) FM synthesis; 4) an LFO; 5) filtered noise

\noindent\textbf{Open-ended Benchmarks}
6) a church bell; 7) a telephone dial tone; 8) a bird call; 9) the sound of waves hitting the ocean; 10) a babbling brook 

These benchmark examples were chosen to test the application of foundational computational sound knowledge in projects that are concise enough for a short LLM-generated code sample to implement. Each of the open-ended benchmarks correspond directly to a specific benchmark in terms of the typical technique used. For example, a church bell sound is often constructed with additive synthesis, and a telephone dial tone is often constructed with AM synthesis.

For each language category, we generated 10 code samples of each benchmark. Thus, we generated 100 code samples total for each language category and 600 code samples total across all our trials. Each of these were manually inspected and evaluated.

\subsection{Prompting}

To prompt the MaxPy, MaxMSP JSON, and Web Audio assistants, we wrote a Python script to call the OpenAI Python library \cite{assistantsAPI}. 
This script creates a new thread for each trial to clear the context, inputted the prompt as the message, parsed the assistant's output for the code and saved it to a new file. 
As Max patches are JSON files with a .maxpat extension, we saved the generated JSON file into a .maxpat file and opened it in Max to evaluate. 
With generated MaxPy scripts, we ran the script, which then produced a .maxpat file that we opened manually and evaluated.
For Web Audio, we ran the generated JavaScript code in our browser to evaluate.

Our Wavir JSON assistant was integrated as a tool on the Wavir website. When a website user clicks on the “Send Prompt” button, the website sends the prompt given to it by the user to the Wavir JSON assistant. 
The assistant then generates a response in JSON which is sent back to the website. 
Wavir then attempts to compile the Wavir JSON, and if it compiles we evaluate the result on screen.

For MaxMSP JSON and MaxPy, we prompted each assistant with the natural language prompt: 
\begin{itemize}
    \item \textit{Based on the examples given, use \{MaxPy, JSON for a Max patch\} to write code that implements \{benchmark\}. }
\end{itemize}
We found that for these languages, adding ``based on the examples given" resulted in far less hallucination, as the assistant would follow the syntax of its knowledge base examples and the guidelines in its knowledge document.

For Web Audio and Wavir JSON, we used this natural language prompt structure: 
\begin{itemize}
    \item \textit{Write \{JSON, Web Audio code\} that implements \{benchmark\}.
}\end{itemize}

This is because we found that for Wavir JSON, contrary to Max JSON and MaxPy, telling the assistant to write code based on the examples hindered the quality of code generated, as all the code generated would just be a copy of the examples. 

For MaxPy Rich Code and Web Audio Rich Code, we added the following line to the end of the corresponding prompt:
\begin{itemize}
    \item \textit{Use for loops and/or \{random(), Math.random()\} in your code.}
\end{itemize}

\subsection{Evaluation Process} \label{evaluation}

\subsubsection{Correctness}

To answer RQ1, we use the \textit{pass@$k$} metric, which measures the probability of success and considers a task successful if any of the $k$ code samples generated pass the test. To evaluate pass@$k$, we generate $n$ samples per task, count the number of correct samples $c \leq n$, and calculate pass@$k$, for some $k$ \cite{codex} as shown below:

\begin{align}
\text{pass@$k$} &:= \mathop{\mathbb{E}}_{\text{Problems}} \left[ 1 - \frac{{\binom{n-c}{k}}} {\binom{n}{k}} \right]
\end{align}

A key difference in our use of pass@$k$ compared to prior work is that we do not have unit tests to measure correctness.
Not only are unit tests difficult to implement for audio DSP across languages, where sample rates and language construct implementations (e.g. low pass filter) may vary, but unit tests are too restrictive for a creative coding task.
For example, it is difficult to formalize a test for ``the sound of waves hitting the ocean''.
Instead, we measure correctness through manual human inspection of the generated solutions.
Due to this labor-intensive evaluation strategy, we limit ourselves to $n=10$ per benchmark and $k \leq 3$ to measure \textbf{pass@1 and pass@3}.

Furthermore, Chen et al. optimize temperature for each value of $k$, since with large values of $k$ (e.g., pass@100) a higher temperature is optimal as the samples generated have more diversity and only one sample need be correct \cite{codex}. However, because we only used small values of $k \leq 3$, we did not find it necessary to tune this hyperparameter for temperature. Instead, we use the default temperature setting of 1, giving us a conservative measure of pass@k - this may be higher with hyperparameter tuning.

Six authors of the paper participated in code evaluation. Code evaluation was conducted in pairs. Each pair generated and evaluated the code for the benchmarks in their assigned languages. 
One pair was assigned MaxMSP JSON, MaxPy, and rich MaxPy, 
one pair was assigned Web Audio and rich Web Audio, and
one pair was assigned Wavir.

Our manual evaluation process was as follows:
First, the evaluator pair was provided with a correct example of each benchmark sound in the target language, coded by a developer on our team. For generated code samples of specific benchmarks, each individual evaluator judges whether the code and the sound it produces resembles the correct example of the benchmark. To be correct, a specific benchmark needs to both produce the sound and have the correct nodes and connections in its code for that particular benchmark. 

For generated code samples of open-ended benchmarks, each individual evaluator judges whether the sound produced resembles, to their ears, the intended sound. In this case, the exact code implementation is less relevant because there are multiple ways to code the sound. There are also many forms an open-ended sound such as a ``bird call" could take while still being considered correct. 

We used the following inter-rater reliability process: if the evaluators agree on pass/fail, then they mark the code sample accordingly. If one or both evaluators are unsure about a sample, or if they disagree, the pair consults as a team consisting of at least one additional evaluator. The team decides through discussion whether to mark the sample correct.

The team repeats this process for each code sample for each benchmark in the language category. The total number of samples marked correct across specific benchmarks, creative benchmarks, and all benchmarks is the variable \textit{c} used to calculate specific, creative, and overall pass@$k$ scores.

\subsubsection{Complexity}

To answer RQ2, we defined complexity as node count. We explain this metric as a measure of complexity in audio DSP in \textit{Threats to Validity}.

For MaxPy and MaxMSP JSON, we wrote a script that parsed the JSON of all the compiled Max patches for a given benchmark in the language and returned the average number of nodes in the JSON.
For Wavir JSON, we counted the number of nodes on screen that the generated Wavir JSON produced if it compiled.
For Web Audio, we counted the number of nodes created in the code (e.g., oscillator and gain nodes), including those added multiple times by loops.

\section{Results}
\label{sec:evaluation}

Here we analyze our results through summary statistics. 
We provide the full evaluation set, with all code generations and raw evaluation scores provided at \\\url{https://github.com/williamyzhang/MaxPy-Wavir-LLM-Generated-Code}

\definecolor{Gray}{gray}{0.5}

\begin{figure}[htbp]
    \centering
    \footnotesize
    \begin{tabular}{|c l|c|c|c|}
        \hline
        \textbf{Language} &  & \textbf{Overall} & \textbf{Specific} & \textbf{Creative} \\
        \hline
        Max (JSON) & & 87 & 44 & 43 \\
        \hline
        MaxPy & & 67 & 43 & 24 \\
        \hline
        MaxPy (Rich Code) & & 48 & 25 & 23 \\
        \hline
        Wavir JSON & & 15 & 9 & 6 \\
        \hline
        Web Audio & & 50 & 30 & 20 \\
        \hline
        Web Audio (Rich Code) & & 56 & 29 & 27 \\
        \hline
    \end{tabular}
    \caption{Number of code samples that are well-formed (parsed/compiled) across all samples overall (100), also split into specific (50) and creative (50) benchmark trials.}
    \label{fig:compiled}
\end{figure}

\definecolor{Gray}{gray}{0.5}

\begin{figure}[htbp]
    \centering
    \footnotesize
    \begin{tabular}{|c l|c|c|c|}
        \hline
        \textbf{Language} &  & \textbf{Overall} & \textbf{Specific} & \textbf{Creative} \\
        \hline
        Max (JSON) & & 30 & 15 & 15 \\
        \hline
        MaxPy & & 31 & 20 & 11 \\
        \hline
        MaxPy (Rich Code) & & 20 & 11 & 9 \\
        \hline
        Wavir JSON & & 8 & 6 & 2 \\
        \hline
        Web Audio & & 38 & 21 & 17 \\
        \hline
        Web Audio (Rich Code) & & 41 & 22 & 19 \\
        \hline
    \end{tabular}
    \caption{Number of code samples that are semantically correct across all, specific, and creative trials.}
    \label{fig:correct}
\end{figure}

\subsection{Correctness}


\subsubsection{MaxMSP}


We calculated the standard pass@$k$ metric for each language across all benchmarks ($n$ = 100). The pass@1 score is 0.300 for MaxMSP JSON, 0.310 for MaxPy normal code, and 0.200 for MaxPy rich code. 
However, the standard pass@$k$ score does not differentiate between semantically incorrect and syntactically incorrect code—a significant difference for LLM-assisted creative coding (see Discussion). To understand the generation better, we separate correctness into two possibilities: well-formedness and semantic correctness.



For a generation to be well-formed the code must run without error. In the case of MaxPy, this means the Python code produces a valid MaxMSP patch. In the case of MaxMSP JSON generation, this means that the JSON produces a valid MaxMSP patch when opened in Max.
Well-formedness is a similar notion to the more commonly used term of syntactical correctness, where code is free of syntactic problems \cite{Corso_2024}. However, well-formedness is a stronger notion than syntactic correctness as code must run to be well-formed, whereas syntactically correct code may still not run. The number of generations for each language that are well-formed is shown in Fig.~\ref{fig:compiled}.

For a generation to be semantically correct, it must correctly implement the benchmark, as judged by our evaluators. 
The number of generations that are semantically correct for each language is shown in Fig.~\ref{fig:correct}. This definition of correctness is used to define $c$ in the calculation of pass@$k$.

\begin{figure}
    \centering
    \includegraphics[width=0.5\textwidth]{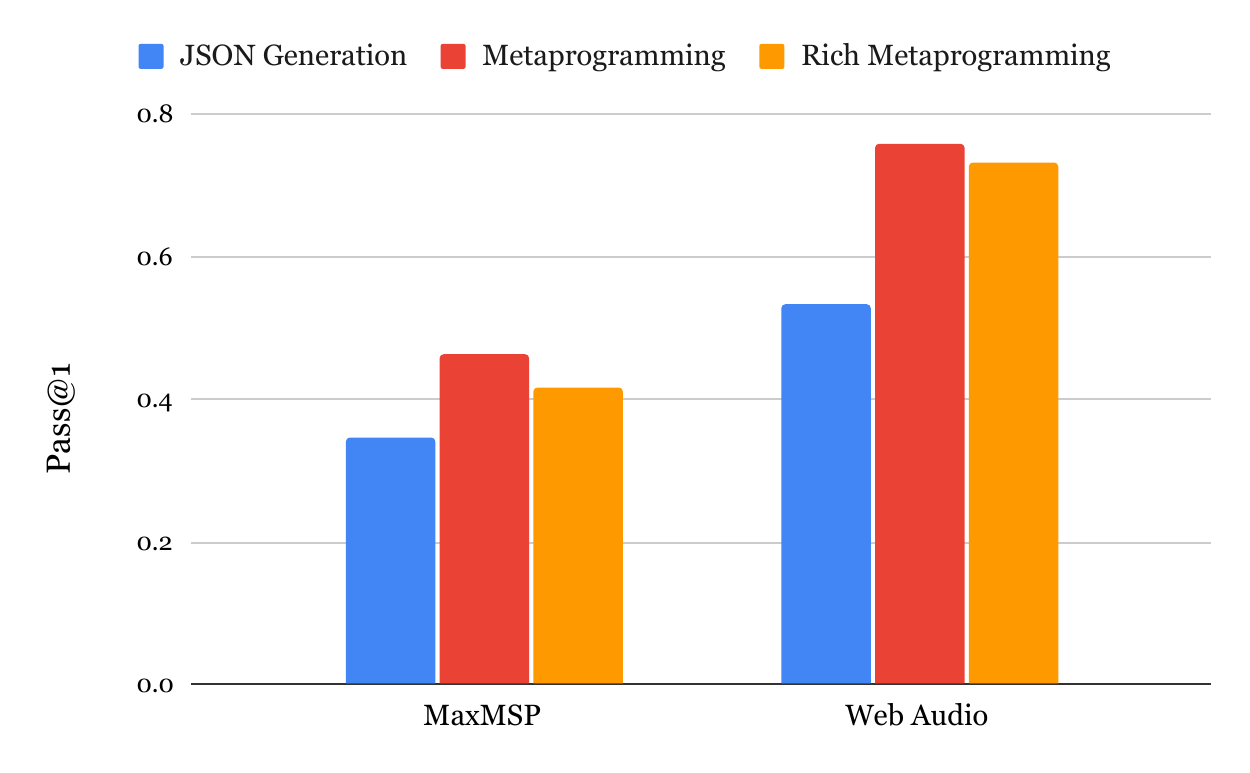}
    \caption{Comparison of pass@1 on well-formed samples across the 3 levels of representation for both sets of languages. Metaprogramming (both normal and rich) performs better than direct JSON generation in both MaxMSP/MaxPy and Wavir/Web Audio.}
    \label{fig:chart-passkcompiled}
\end{figure}


We calculated pass@$k$ scores again where $n$ was set to the number of well-formed code samples for each language, rather than the total number of samples generated.
Here, the \textbf{pass@1 score is 0.345 for MaxMSP JSON, 0.463 for MaxPy normal code, and 0.417 for MaxPy rich code}.

For the MaxMSP languages, when we calculate pass@$k$ as semantic correctness over all code samples, it is unclear whether metaprogramming or direct JSON generation results in more correct code, while rich metaprogramming performs worse than both normal metaprogramming and direct JSON generation.

However, when we calculate pass@$k$ as semantic correctness over only well-formed samples, both normal and rich metaprogramming result in higher rates of correctness over direct JSON generation. Rich metaprogramming still performs worse than normal metaprogramming.

\subsubsection{Web Audio}


Over all code samples ($n$ = 100), the pass@1 score is 0.080 for Wavir JSON, 0.380 for Web Audio normal code, and 0.410 for Web Audio rich code. 

Calculating pass@$k$ over only the well-formed samples, the \textbf{pass@1 score is 0.533 for Wavir JSON, 0.760 for Web Audio normal code, and 0.732 for Web Audio rich code}. 

For the Web Audio languages, metaprogramming results in higher correctness than direct JSON generation across both pass@$k$ metrics. Rich metaprogramming performs either better or worse depending on the metric used.

\subsubsection{Conclusions}

\textbf{To answer RQ1, metaprogramming and rich metaprogramming result in more correct code than direct JSON generation when we measure semantic correctness over only well-formed samples.}  Results are less conclusive when using a standard pass@$k$ metric of semantic correctness over all generated samples. We will explain why calculating pass@$k$ as semantic correctness over well-formed samples is a useful metric and how the research community can use these results in the Discussion.

\subsection{Complexity}

\definecolor{Gray}{gray}{0.5}
\begin{figure}[htbp]
    \centering
    \footnotesize
    \begin{tabular}{|c l|c|c|c|}
        \hline
        \textbf{Language} &  & \textbf{Overall} & \textbf{Specific} & \textbf{Creative} \\
        \hline
        MaxMSP (JSON) & &
        5.1 &
        5.2 &
        5.0 \\
        \hline
        MaxPy
        &  &  6.7 &  5.4 &  8.0 \\
        \hline
        
        MaxPy (Rich Code)
        &  &  14.4 &  12.3 &  16.5 \\
        \hline

        Wavir (JSON)
        &  &  4.6 &  4.7 &  4.3 \\
        \hline

        Web Audio
        &  &  4.3 &  4.5 &  4.1 \\
        \hline

        Web Audio (Rich Code)
        &  &  9.0 &  7.6 &  10.3 \\
        \hline
    \end{tabular}
    \caption{Average complexity (number of nodes) in well-formed samples. }
    \label{fig:complexity}
\end{figure}

Complexity was measured as the average number of nodes per compiled code sample for each benchmark. The significance of the difference in complexity between each pair of languages was calculated using a one-sided Wilcoxon signed-rank test across the averages for each benchmark. 

MaxPy Rich Code had a statistically significantly higher node count than either MaxMSP JSON or normal MaxPy, with p-values of 0.002 and 0.001 respectively. 
Web Audio Rich Code also had a statistically significantly higher average node count than either Wavir JSON or normal Web Audio, with p-values of 0.02 and 0.01 respectively.

Normal MaxPy had a statistically significantly higher average than MaxMSP JSON at p = 0.05. However, Wavir JSON had a slightly higher average node count than Web Audio, but the difference was not statistically significant.

\subsubsection{Conclusions}
\textbf{To answer RQ2, rich metaprogramming as the target of LLM generation results in the most complex code.} We found no consistent difference between normal metaprogramming and JSON generation in complexity.

\section{Discussion} \label{discussion}

Here we explain our new correctness metric, the applications of our results, and potential limitations of our study.

\subsection{A New Correctness Metric}

We calculated two different pass@$k$ metrics in our evaluation, and we derived our conclusion that metaprogramming results in more semantically correct code given it is well-formed from the second metric (semantically correct over only well-formed samples). Here, we will justify that this second metric is a reasonable measurement of correctness with practical implications.

Most evaluations of LLM-generated code rely on a correctness metric that considers code as correct if it passes the unit test, and incorrect otherwise \cite{codex}. Under this metric, semantic errors and syntax errors are treated the same. However, distinguishing between these two types of errors is significant for AI-based coding assistants. 
Corso et al. investigated both syntactic and semantic correctness of AI-generated code, concluding that a simple pass-fail check for correctness using test cases is insufficient, as AI assistants may write code that passes unit tests but does not properly implement the function requested \cite{Corso_2024}.

An important concern with LLM-based coding assistants in the creative coding and visual language domain is their accessibility to non-programmers. For non-programmers, if an LLM-generated code sample is not producing the result desired, it is likely easier to regenerate another code sample rather than debug the broken code. A tool that is fully accessible to non-programmers would ideally not require directly debugging even a high-level programming language \cite{ferdowsi2023coldeco}.

When building an LLM-based coding assistant, it is easy to automate a check for well-formedness, as we only need to check for errors in the console. On the other hand, checking for semantic correctness in creative coding is far more difficult and subjective to some extent. The latter would require a unit test for whether a sound project coded by the AI assistant sounds convincingly like a bird call, and if it is the exact type of bird call the user desires. Thus, a hypothetical assistant can filter out samples that are not well-formed and return only well-formed samples to the user. It is important that the generated well-formed code has a high rate of semantic correctness. Therefore, semantic correctness over only well-formed samples is a practical metric for such an assistant. 

\textbf{In designing a creative coding assistant accessible to non-programmers, prioritizing code generation that yields semantically correct code, given that the generation is well-formed, is more useful than prioritizing semantically correct code over all generations.}

\subsection{Metaprogramming as the Target of LLM Code Generation}

As Wavir and MaxMSP are designed as visual languages in order to give users a high-level interface to computation, effective use of LLM-assisted tools that generate visual code could help further raise the level of abstraction for users. However, generating code (rather than a blackbox system), is important so that users maintain control over their output. 

Metaprogramming offers concision and efficiency over JSON in representing a visual language. To generate the same MaxMSP patch requires far more lines of code in JSON than in MaxPy. Moreover, metaprogramming is much more intuitive than JSON for organizing and connecting elements of the visual language, as it encodes the visual language in a high-level programming language instead of lower-level JSON data. For these reasons, metaprogramming is certainly preferable for human programmers when choosing a text-based representation of visual languages. 

Our results indicate that generated code for visual dataflow languages is more semantically correct at the metaprogramming level than in JSON representation, given it is well-formed. This suggests that metaprogramming is preferable for LLM-based coding assistants as well. This preference has important implications for LLM code generation. \textbf{When the target of generation has multiple representations in text, a higher-level representation (i.e., more abstract, closer to natural language) may be more semantically correct, given the generation is well-formed, than lower-level data representations.}

\subsection{Benefits of Complexity}

The goal of measuring code complexity is most often to limit the complexity of programs, since more complex programs are harder to understand, more error-prone, and more difficult to debug \cite{Corso_2024, lavazza2023understandability}. 
However, here we will argue for some advantages that complexity offers in creative coding.

With regard to creativity in computational systems, Wiggins formalizes the concept of a ``perfect or productive aberration", where new concepts outside of the existing domain produce valuable results \cite{wiggins2006searching, wiggins2006framework}. Applying this concept to creative coding, a productive aberration would result when a programmer uses the language in an unexpected way outside of their knowledge domain yet generates a useful or pleasing outcome. Much as productive aberrations in computational agents are useful for training and improving an AI \cite{wiggins2006framework}, productive aberrations in creative coding allow the programmer to develop in knowledge and mastery of the language.

Following the notion of productive aberrations, an LLM-based coding assistant that can offer more complexity would allow creative coders to use the language in ways outside of their knowledge. Specific to our domain, an assistant capable of generating programs with more nodes is more likely to result in unexpected outcomes or usages of the language that lead to productive aberrations.

A historical example of the advantages of complexity is FM synthesis. In describing sound synthesis via frequency modulation, Chowning emphasized the complexity of the components involved in FM synthesis as a major contributor to the rich sounds produced by the technique \cite{chowning1977complexity}. These complex sounds subsequently made FM central to the synth-dominated music of the 1980s. 

Of course, excess complexity will be hard to understand and inaccessible to beginners, so an LLM-based coding assistant that can generate complex code should also be able to generate simpler code if requested. For this purpose, the results of our study are relevant. Rich metaprogramming resulted in the most complex code, whereas the complexity of normal metaprogramming and direct JSON generation was approximately equal. Thus, metaprogramming can provide both a low level of complexity similar to that of JSON generation and a high level of complexity via prompting for rich code. The availability of multiple modes is an inherent advantage of metaprogramming over direct JSON generation, which lacks features such as loops and randomness that result in rich code.

\textbf{Rich metaprogramming allows an LLM-based coding assistant to provide users with the creative exploration and development afforded by complexity. Meanwhile, switching between rich and normal modes of metaprogramming code generation would allow an LLM-based coding assistant to offer more complex or more basic solutions upon request.}

\subsection{Threats to Validity}

\begin{figure}
    \centering
    \includegraphics[width=0.49\textwidth]{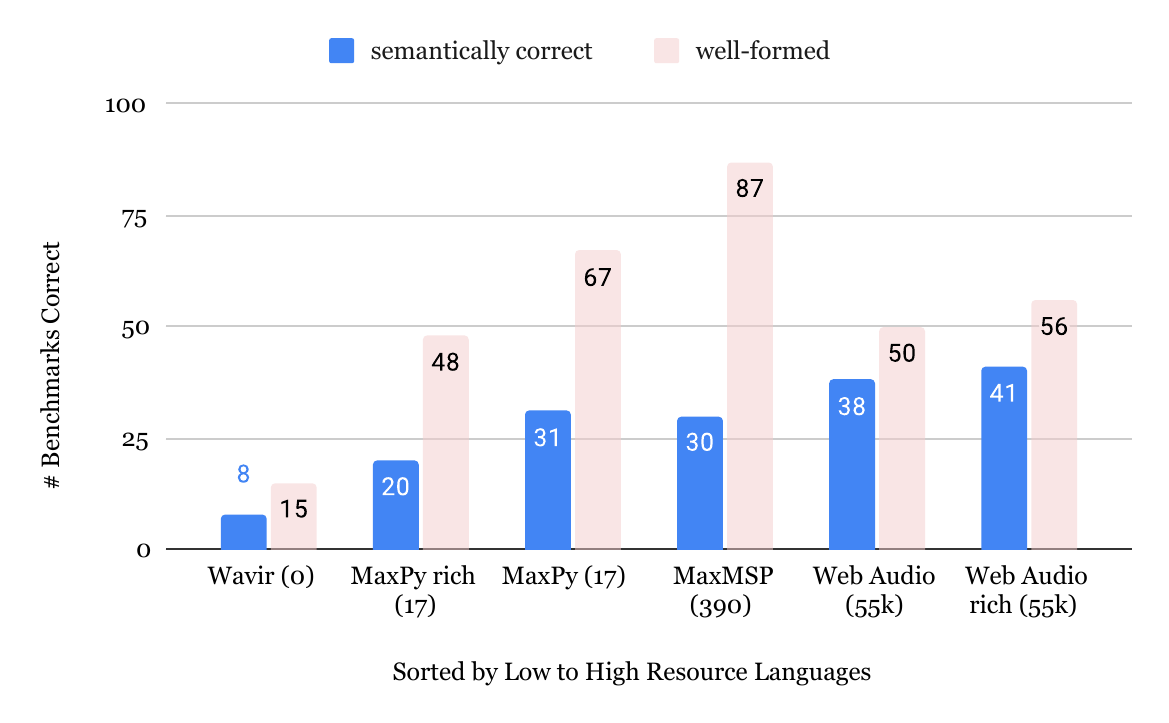}
    \caption{Number of correct benchmark generations for both well-formedness and semantic correctness. Each language is notated with the number of code examples in this language found on GitHub.}
    \label{fig:overview}
\end{figure}

\subsubsection{Confounding Variables for Correctness}

The major internal threat to the validity of our study is the conflation of correctness because of the nature of the language with correctness because of the LLM having better training data. 

Programming languages with more training data in the LLM's training corpus tend to perform better in evaluations of generated code, as defined by pass@$k$ \cite{athiwaratkun2023multilingual, cassano2022multiple, cassano2024lowresource}. GitHub is frequently used as a source of code examples for training LLMs on code \cite{kocetkov2022stack3tb}. With closed-source LLMs such as GPT-4, the training data used is not released to the public \cite{openai2024gpt4technicalreport, kocetkov2022stack3tb}. However, even with closed-source models such as Codex, a language's popularity on GitHub is a predictor of its performance \cite{cassano2022multiple}. Therefore, we classify the languages used in our paper from low- to high-resource languages based on the number of code examples available on GitHub.

As Fig.~\ref{fig:overview} shows, the number of semantically correct generations tends to increase for higher resource languages.
As high-resource languages likely have far more representation in GPT-4's training corpus, the pass@$k$ scores for high-resource languages such as Web Audio may be higher due to their greater available training data, rather than their characteristics as a language. MaxPy's better performance over MaxMSP in the metric of semantic correctness given syntactic correctness shows that higher correctness is not solely due to the amount of training data. Nonetheless, future research should control for the training data disparity between metaprogramming languages and direct JSON inputs.

\subsubsection{Complexity Metric} \label{threats to validity - complexity}

While we used node count as our complexity metric, there are other more commonly used metrics for code complexity, such as cyclomatic complexity. In this section we justify our choice of node count over other metrics.

Cyclomatic complexity, as first defined by McCabe, measures complexity as the number of linearly independent paths through a program \cite{mccabe1976complexity}. However, cyclomatic complexity does not make sense in the audio DSP programs used in our study because these programs lack control flow. That is, there is only one path through the programs, since they are designed to generate one sound in one way, without branching into multiple possibilities. Without control flow, their cyclomatic complexity will be the same—1. Therefore, cyclomatic complexity is unfit as a measure of complexity for our study.

Another classic measure of complexity is lines of code \cite{lavazza2023understandability}. Lines of code has been shown to perform approximately as well as other metrics, including cyclomatic complexity, in predicting empirical measures of code understandability \cite{lavazza2023understandability,davis1988complexity}. As our study used multiple levels of representation of visual programming languages, however, measuring lines of code would not allow for comparison between the different languages across these levels of representation. Nonetheless, the logic behind lines of code is useful in our study, measuring complexity as the number of elements in a program \cite{davis1988complexity}.

In the context of visual node-based programming languages, a node is a base element of the program. The number of nodes can be a visual dataflow analog of lines of code for traditional coding languages. 
In fact, node count likely underestimates complexity, since with increased number of nodes there is also an exponential increase in the interactions between nodes. A program with 4 nodes is usually more than twice as complex as a program with only 2 nodes. Complexity scales more than linearly with the number of nodes. Therefore, node count is a conservative estimate of a lower bound on complexity.

\section*{Conclusion}

Node-based visual dataflow languages offer an accessible form of programming to users with limited coding knowledge and experience. LLM code generation presents an opportunity to further lower the barrier to using programming languages. In this paper, we contribute to the knowledge in these domains by determining the best strategy of LLM code generation for visual languages. We evaluate code generated by LLM-based assistants for two visual node-based languages for audio programming across three levels of code representation: JSON, metaprogramming, and rich metaprogramming. Our results indicate that metaprogramming code representations result in more semantically correct code, given that the code is well-formed. Our results also show that rich metaprogramming results in higher code complexity. The finding of metaprogramming's higher semantic correctness when it is well-formed may be useful to future work in researching and building LLM-based coding assistants, as it suggests a higher and more abstract level of representation of the target generation is preferable. 

\bibliography{refs}

\begin{thebibliography}{31}
\providecommand{\natexlab}[1]{#1}

\bibitem[{wav(2024)}]{wavir}
 2024.
\newblock Wavir.
\newblock \url{https://wavir.io/}.

\bibitem[{Angert et~al.(2023)Angert, Suzara, Han, Pondoc, and Subramonyam}]{spellburst}
Angert, T.; Suzara, M.~I.; Han, J.; Pondoc, C.~L.; and Subramonyam, H. 2023.
\newblock {Spellburst: A node-based Interface for Exploratory Creative Coding with Natural Language Prompts}.

\bibitem[{Athiwaratkun et~al.(2023)Athiwaratkun, Gouda, Wang, Li, Tian, Tan, Ahmad, Wang, Sun, Shang, Gonugondla, Ding, Kumar, Fulton, Farahani, Jain, Giaquinto, Qian, Ramanathan, Nallapati, Ray, Bhatia, Sengupta, Roth, and Xiang}]{athiwaratkun2023multilingual}
Athiwaratkun, B.; Gouda, S.~K.; Wang, Z.; Li, X.; Tian, Y.; Tan, M.; Ahmad, W.~U.; Wang, S.; Sun, Q.; Shang, M.; Gonugondla, S.~K.; Ding, H.; Kumar, V.; Fulton, N.; Farahani, A.; Jain, S.; Giaquinto, R.; Qian, H.; Ramanathan, M.~K.; Nallapati, R.; Ray, B.; Bhatia, P.; Sengupta, S.; Roth, D.; and Xiang, B. 2023.
\newblock Multi-lingual Evaluation of Code Generation Models.
\newblock arXiv:2210.14868.

\bibitem[{Cassano et~al.(2024)Cassano, Gouwar, Lucchetti, Schlesinger, Freeman, Anderson, Feldman, Greenberg, Jangda, and Guha}]{cassano2024lowresource}
Cassano, F.; Gouwar, J.; Lucchetti, F.; Schlesinger, C.; Freeman, A.; Anderson, C.~J.; Feldman, M.~Q.; Greenberg, M.; Jangda, A.; and Guha, A. 2024.
\newblock Knowledge Transfer from High-Resource to Low-Resource Programming Languages for Code LLMs.
\newblock arXiv:2308.09895.

\bibitem[{Cassano et~al.(2022)Cassano, Gouwar, Nguyen, Nguyen, Phipps-Costin, Pinckney, Yee, Zi, Anderson, Feldman, Guha, Greenberg, and Jangda}]{cassano2022multiple}
Cassano, F.; Gouwar, J.; Nguyen, D.; Nguyen, S.; Phipps-Costin, L.; Pinckney, D.; Yee, M.-H.; Zi, Y.; Anderson, C.~J.; Feldman, M.~Q.; Guha, A.; Greenberg, M.; and Jangda, A. 2022.
\newblock MultiPL-E: A Scalable and Extensible Approach to Benchmarking Neural Code Generation.
\newblock arXiv:2208.08227.

\bibitem[{Celani and Vaz(2012)}]{grasshopper}
Celani, G.; and Vaz, C. E.~V. 2012.
\newblock CAD Scripting and Visual Programming Languages for Implementing Computational Design Concepts: A Comparison from a Pedagogical Point of View.
\newblock \emph{International Journal of Architectural Computing}, 10(1): 121--137.

\bibitem[{Chen et~al.(2021)Chen, Tworek, Jun, Yuan, Ponde, Kaplan, Edwards, Burda, Joseph, Brockman, Ray, Puri, Krueger, Petrov, Khlaaf, Sastry, Mishkin, Chan, Gray, Ryder, Pavlov, Power, Kaiser, Bavarian, Winter, Tillet, Such, Cummings, Plappert, Chantzis, Barnes, Herbert-Voss, Guss, Nichol, Babuschkin, Balaji, Jain, Carr, Leike, Achiam, Misra, Morikawa, Radford, Knight, Brundage, Murati, Mayer, Welinder, McGrew, Amodei, McCandlish, Sutskever, and Zaremba}]{codex}
Chen, M.; Tworek, J.; Jun, H.; Yuan, Q.; Ponde, H.; Kaplan, J.; Edwards, H.; Burda, Y.; Joseph, N.; Brockman, G.; Ray, A.; Puri, R.; Krueger, G.; Petrov, M.; Khlaaf, H.; Sastry, G.; Mishkin, P.; Chan, B.; Gray, S.; Ryder, N.; Pavlov, M.; Power, A.; Kaiser, L.; Bavarian, M.; Winter, C.; Tillet, P.; Such, F.; Cummings, D.; Plappert, M.; Chantzis, F.; Barnes, E.; Herbert-Voss, A.; Guss, W.; Nichol, A.; Babuschkin, I.; Balaji, S.; Jain, S.; Carr, A.; Leike, J.; Achiam, J.; Misra, V.; Morikawa, E.; Radford, A.; Knight, M.; Brundage, M.; Murati, M.; Mayer, K.; Welinder, P.; McGrew, B.; Amodei, D.; McCandlish, S.; Sutskever, I.; and Zaremba, W. 2021.
\newblock {Evaluating Large Language Models Trained on Code}.
\newblock arXiv:2107.03374.

\bibitem[{Chowning(1977)}]{chowning1977complexity}
Chowning, J.~M. 1977.
\newblock The Synthesis of Complex Audio Spectra by Means of Frequency Modulation.
\newblock \emph{Computer Music Journal}, 1(2): 46--54.

\bibitem[{Corso et~al.(2024)Corso, Mariani, Micucci, and Riganelli}]{Corso_2024}
Corso, V.; Mariani, L.; Micucci, D.; and Riganelli, O. 2024.
\newblock Generating Java Methods: An Empirical Assessment of Four AI-Based Code Assistants.
\newblock In \emph{Proceedings of the 32nd IEEE/ACM International Conference on Program Comprehension}, ICPC ’24. ACM.

\bibitem[{{Cycling '74}(2023)}]{maxmsp}
{Cycling '74}. 2023.
\newblock {MaxMSP product page}.
\newblock \url{https://cycling74.com/products/max}.

\bibitem[{Davis and LeBlanc(1988)}]{davis1988complexity}
Davis, J.; and LeBlanc, R. 1988.
\newblock A study of the applicability of complexity measures.
\newblock \emph{IEEE Transactions on Software Engineering}, 14(9): 1366--1372.

\bibitem[{Ferdowsi et~al.(2023)Ferdowsi, Williams, Drosos, Gordon, Negreanu, Polikarpova, Sarkar, and Zorn}]{ferdowsi2023coldeco}
Ferdowsi, K.; Williams, J.; Drosos, I.; Gordon, A.~D.; Negreanu, C.; Polikarpova, N.; Sarkar, A.; and Zorn, B. 2023.
\newblock ColDeco: An end user spreadsheet inspection tool for AI-generated code.
\newblock In \emph{2023 IEEE Symposium on Visual Languages and Human-Centric Computing (VL/HCC)}, 82--91. IEEE.

\bibitem[{Hempel, Lubin, and Chugh(2019)}]{hempel2019sketch}
Hempel, B.; Lubin, J.; and Chugh, R. 2019.
\newblock Sketch-n-sketch: Output-directed programming for svg.
\newblock In \emph{Proceedings of the 32nd Annual ACM Symposium on User Interface Software and Technology}, 281--292.

\bibitem[{Jensen et~al.(2007)Jensen, Francis, Larsen, and Christensen}]{shadergraphs}
Jensen, P. D.~E.; Francis, N.; Larsen, B.~D.; and Christensen, N.~J. 2007.
\newblock Interactive shader development.
\newblock In \emph{Proceedings of the 2007 ACM SIGGRAPH Symposium on Video Games}, Sandbox '07, 89–95. New York, NY, USA: Association for Computing Machinery.
\newblock ISBN 9781595937490.

\bibitem[{Jimenez et~al.(2023)Jimenez, Yang, Wettig, Yao, Pei, Press, and Narasimhan}]{jimenez2023swe}
Jimenez, C.~E.; Yang, J.; Wettig, A.; Yao, S.; Pei, K.; Press, O.; and Narasimhan, K. 2023.
\newblock Swe-bench: Can language models resolve real-world github issues?
\newblock \emph{arXiv preprint arXiv:2310.06770}.

\bibitem[{Kocetkov et~al.(2022)Kocetkov, Li, Allal, Li, Mou, Ferrandis, Jernite, Mitchell, Hughes, Wolf, Bahdanau, von Werra, and de~Vries}]{kocetkov2022stack3tb}
Kocetkov, D.; Li, R.; Allal, L.~B.; Li, J.; Mou, C.; Ferrandis, C.~M.; Jernite, Y.; Mitchell, M.; Hughes, S.; Wolf, T.; Bahdanau, D.; von Werra, L.; and de~Vries, H. 2022.
\newblock The Stack: 3 TB of permissively licensed source code.
\newblock arXiv:2211.15533.

\bibitem[{Lavazza, Morasca, and Gatto(2023)}]{lavazza2023understandability}
Lavazza, L.; Morasca, S.; and Gatto, M. 2023.
\newblock An empirical study on software understandability and its dependence on code characteristics.
\newblock \emph{Empirical Softw. Engg.}, 28(6).

\bibitem[{Lin and Martelaro(2023)}]{lin2023jigsaw}
Lin, D. C.-E.; and Martelaro, N. 2023.
\newblock Jigsaw: Supporting Designers in Prototyping Multimodal Applications by Assembling AI Foundation Models.
\newblock \emph{arXiv preprint arXiv:2310.08574}.

\bibitem[{Liu et~al.(2023{\natexlab{a}})Liu, Xia, Wang, and Zhang}]{isCodeCorrect}
Liu, J.; Xia, C.; Wang, Y.; and Zhang, L. 2023{\natexlab{a}}.
\newblock {Is Your Code Generated by ChatGPT Really Correct? Rigorous Evaluation of Large Language Models for Code Generation}.

\bibitem[{Liu et~al.(2023{\natexlab{b}})Liu, Peterson, Lee, and Santolucito}]{maxpy}
Liu, R.; Peterson, S.; Lee, R.; and Santolucito, M. 2023{\natexlab{b}}.
\newblock {Maxpy: An open-source Python package for programmatic construction and manipulation of MaxMSP patches}.
\newblock Available at \url{https://github.com/Barnard-PL-Labs/MaxPy/blob/main/MaxPy-NIME-2023-Paper.pdf}.

\bibitem[{McCabe(1976)}]{mccabe1976complexity}
McCabe, T. 1976.
\newblock A Complexity Measure.
\newblock \emph{IEEE Transactions on Software Engineering}, SE-2(4): 308--320.

\bibitem[{McNutt et~al.(2023)McNutt, Wang, Deline, and Drucker}]{mcnuttCHI2023}
McNutt, A.~M.; Wang, C.; Deline, R.~A.; and Drucker, S.~M. 2023.
\newblock On the Design of AI-powered Code Assistants for Notebooks.
\newblock In \emph{Proceedings of the 2023 CHI Conference on Human Factors in Computing Systems}, CHI '23. New York, NY, USA: Association for Computing Machinery.
\newblock ISBN 9781450394215.

\bibitem[{{OpenAI}(2024)}]{assistantsAPI}
{OpenAI}. 2024.
\newblock Assistants API.

\bibitem[{OpenAI et~al.(2024)OpenAI, Achiam, Adler, Agarwal, Ahmad, Akkaya, Aleman, Almeida, Altenschmidt, Altman, Anadkat, Avila, Babuschkin, Balaji, Balcom, Baltescu, Bao, Bavarian, Belgum, Bello, Berdine, Bernadett-Shapiro, Berner, Bogdonoff, Boiko, Boyd, Brakman, Brockman, Brooks, Brundage, Button, Cai, Campbell, Cann, Carey, Carlson, Carmichael, Chan, Chang, Chantzis, Chen, Chen, Chen, Chen, Chen, Chess, Cho, Chu, Chung, Cummings, Currier, Dai, Decareaux, Degry, Deutsch, Deville, Dhar, Dohan, Dowling, Dunning, Ecoffet, Eleti, Eloundou, Farhi, Fedus, Felix, Fishman, Forte, Fulford, Gao, Georges, Gibson, Goel, Gogineni, Goh, Gontijo-Lopes, Gordon, Grafstein, Gray, Greene, Gross, Gu, Guo, Hallacy, Han, Harris, He, Heaton, Heidecke, Hesse, Hickey, Hickey, Hoeschele, Houghton, Hsu, Hu, Hu, Huizinga, Jain, Jain, Jang, Jiang, Jiang, Jin, Jin, Jomoto, Jonn, Jun, Kaftan, Łukasz Kaiser, Kamali, Kanitscheider, Keskar, Khan, Kilpatrick, Kim, Kim, Kim, Kirchner, Kiros, Knight, Kokotajlo, Łukasz Kondraciuk,
  Kondrich, Konstantinidis, Kosic, Krueger, Kuo, Lampe, Lan, Lee, Leike, Leung, Levy, Li, Lim, Lin, Lin, Litwin, Lopez, Lowe, Lue, Makanju, Malfacini, Manning, Markov, Markovski, Martin, Mayer, Mayne, McGrew, McKinney, McLeavey, McMillan, McNeil, Medina, Mehta, Menick, Metz, Mishchenko, Mishkin, Monaco, Morikawa, Mossing, Mu, Murati, Murk, Mély, Nair, Nakano, Nayak, Neelakantan, Ngo, Noh, Ouyang, O'Keefe, Pachocki, Paino, Palermo, Pantuliano, Parascandolo, Parish, Parparita, Passos, Pavlov, Peng, Perelman, de~Avila Belbute~Peres, Petrov, de~Oliveira~Pinto, Michael, Pokorny, Pokrass, Pong, Powell, Power, Power, Proehl, Puri, Radford, Rae, Ramesh, Raymond, Real, Rimbach, Ross, Rotsted, Roussez, Ryder, Saltarelli, Sanders, Santurkar, Sastry, Schmidt, Schnurr, Schulman, Selsam, Sheppard, Sherbakov, Shieh, Shoker, Shyam, Sidor, Sigler, Simens, Sitkin, Slama, Sohl, Sokolowsky, Song, Staudacher, Such, Summers, Sutskever, Tang, Tezak, Thompson, Tillet, Tootoonchian, Tseng, Tuggle, Turley, Tworek, Uribe, Vallone,
  Vijayvergiya, Voss, Wainwright, Wang, Wang, Wang, Ward, Wei, Weinmann, Welihinda, Welinder, Weng, Weng, Wiethoff, Willner, Winter, Wolrich, Wong, Workman, Wu, Wu, Wu, Xiao, Xu, Yoo, Yu, Yuan, Zaremba, Zellers, Zhang, Zhang, Zhao, Zheng, Zhuang, Zhuk, and Zoph}]{openai2024gpt4technicalreport}
OpenAI; Achiam, J.; Adler, S.; Agarwal, S.; Ahmad, L.; Akkaya, I.; Aleman, F.~L.; Almeida, D.; Altenschmidt, J.; Altman, S.; Anadkat, S.; Avila, R.; Babuschkin, I.; Balaji, S.; Balcom, V.; Baltescu, P.; Bao, H.; Bavarian, M.; Belgum, J.; Bello, I.; Berdine, J.; Bernadett-Shapiro, G.; Berner, C.; Bogdonoff, L.; Boiko, O.; Boyd, M.; Brakman, A.-L.; Brockman, G.; Brooks, T.; Brundage, M.; Button, K.; Cai, T.; Campbell, R.; Cann, A.; Carey, B.; Carlson, C.; Carmichael, R.; Chan, B.; Chang, C.; Chantzis, F.; Chen, D.; Chen, S.; Chen, R.; Chen, J.; Chen, M.; Chess, B.; Cho, C.; Chu, C.; Chung, H.~W.; Cummings, D.; Currier, J.; Dai, Y.; Decareaux, C.; Degry, T.; Deutsch, N.; Deville, D.; Dhar, A.; Dohan, D.; Dowling, S.; Dunning, S.; Ecoffet, A.; Eleti, A.; Eloundou, T.; Farhi, D.; Fedus, L.; Felix, N.; Fishman, S.~P.; Forte, J.; Fulford, I.; Gao, L.; Georges, E.; Gibson, C.; Goel, V.; Gogineni, T.; Goh, G.; Gontijo-Lopes, R.; Gordon, J.; Grafstein, M.; Gray, S.; Greene, R.; Gross, J.; Gu, S.~S.; Guo, Y.; Hallacy,
  C.; Han, J.; Harris, J.; He, Y.; Heaton, M.; Heidecke, J.; Hesse, C.; Hickey, A.; Hickey, W.; Hoeschele, P.; Houghton, B.; Hsu, K.; Hu, S.; Hu, X.; Huizinga, J.; Jain, S.; Jain, S.; Jang, J.; Jiang, A.; Jiang, R.; Jin, H.; Jin, D.; Jomoto, S.; Jonn, B.; Jun, H.; Kaftan, T.; Łukasz Kaiser; Kamali, A.; Kanitscheider, I.; Keskar, N.~S.; Khan, T.; Kilpatrick, L.; Kim, J.~W.; Kim, C.; Kim, Y.; Kirchner, J.~H.; Kiros, J.; Knight, M.; Kokotajlo, D.; Łukasz Kondraciuk; Kondrich, A.; Konstantinidis, A.; Kosic, K.; Krueger, G.; Kuo, V.; Lampe, M.; Lan, I.; Lee, T.; Leike, J.; Leung, J.; Levy, D.; Li, C.~M.; Lim, R.; Lin, M.; Lin, S.; Litwin, M.; Lopez, T.; Lowe, R.; Lue, P.; Makanju, A.; Malfacini, K.; Manning, S.; Markov, T.; Markovski, Y.; Martin, B.; Mayer, K.; Mayne, A.; McGrew, B.; McKinney, S.~M.; McLeavey, C.; McMillan, P.; McNeil, J.; Medina, D.; Mehta, A.; Menick, J.; Metz, L.; Mishchenko, A.; Mishkin, P.; Monaco, V.; Morikawa, E.; Mossing, D.; Mu, T.; Murati, M.; Murk, O.; Mély, D.; Nair, A.; Nakano, R.;
  Nayak, R.; Neelakantan, A.; Ngo, R.; Noh, H.; Ouyang, L.; O'Keefe, C.; Pachocki, J.; Paino, A.; Palermo, J.; Pantuliano, A.; Parascandolo, G.; Parish, J.; Parparita, E.; Passos, A.; Pavlov, M.; Peng, A.; Perelman, A.; de~Avila Belbute~Peres, F.; Petrov, M.; de~Oliveira~Pinto, H.~P.; Michael; Pokorny; Pokrass, M.; Pong, V.~H.; Powell, T.; Power, A.; Power, B.; Proehl, E.; Puri, R.; Radford, A.; Rae, J.; Ramesh, A.; Raymond, C.; Real, F.; Rimbach, K.; Ross, C.; Rotsted, B.; Roussez, H.; Ryder, N.; Saltarelli, M.; Sanders, T.; Santurkar, S.; Sastry, G.; Schmidt, H.; Schnurr, D.; Schulman, J.; Selsam, D.; Sheppard, K.; Sherbakov, T.; Shieh, J.; Shoker, S.; Shyam, P.; Sidor, S.; Sigler, E.; Simens, M.; Sitkin, J.; Slama, K.; Sohl, I.; Sokolowsky, B.; Song, Y.; Staudacher, N.; Such, F.~P.; Summers, N.; Sutskever, I.; Tang, J.; Tezak, N.; Thompson, M.~B.; Tillet, P.; Tootoonchian, A.; Tseng, E.; Tuggle, P.; Turley, N.; Tworek, J.; Uribe, J. F.~C.; Vallone, A.; Vijayvergiya, A.; Voss, C.; Wainwright, C.; Wang,
  J.~J.; Wang, A.; Wang, B.; Ward, J.; Wei, J.; Weinmann, C.; Welihinda, A.; Welinder, P.; Weng, J.; Weng, L.; Wiethoff, M.; Willner, D.; Winter, C.; Wolrich, S.; Wong, H.; Workman, L.; Wu, S.; Wu, J.; Wu, M.; Xiao, K.; Xu, T.; Yoo, S.; Yu, K.; Yuan, Q.; Zaremba, W.; Zellers, R.; Zhang, C.; Zhang, M.; Zhao, S.; Zheng, T.; Zhuang, J.; Zhuk, W.; and Zoph, B. 2024.
\newblock GPT-4 Technical Report.
\newblock arXiv:2303.08774.

\bibitem[{Santolucito et~al.(2018{\natexlab{a}})Santolucito, Goldman, Weseley, and Piskac}]{santolucito2018programming}
Santolucito, M.; Goldman, D.; Weseley, A.; and Piskac, R. 2018{\natexlab{a}}.
\newblock Programming by Example: Efficient, but Not" Helpful".
\newblock In \emph{9th Workshop on Evaluation and Usability of Programming Languages and Tools (PLATEAU 2018)}.

\bibitem[{Santolucito, Hallahan, and Piskac(2019)}]{santolucito2019live}
Santolucito, M.; Hallahan, W.~T.; and Piskac, R. 2019.
\newblock Live programming by example.
\newblock In \emph{Extended abstracts of the 2019 CHI conference on human factors in computing systems}, 1--4.

\bibitem[{Santolucito et~al.(2018{\natexlab{b}})Santolucito, Rogers, Lombardo, and Piskac}]{santolucito2018pbe}
Santolucito, M.; Rogers, K.; Lombardo, A.; and Piskac, R. 2018{\natexlab{b}}.
\newblock Programming-by-example for audio: synthesizing digital signal processing programs.
\newblock In \emph{Proceedings of the 6th ACM SIGPLAN International Workshop on Functional Art, Music, Modeling, and Design}, 18--25.

\bibitem[{Wang et~al.(2024)Wang, Yin, Hu, Mao, and Hui}]{wang2024exploring}
Wang, A.; Yin, Z.; Hu, Y.; Mao, Y.; and Hui, P. 2024.
\newblock Exploring the Potential of Large Language Models in Artistic Creation: Collaboration and Reflection on Creative Programming.
\newblock \emph{arXiv preprint arXiv:2402.09750}.

\bibitem[{Wiggins(2006{\natexlab{a}})}]{wiggins2006framework}
Wiggins, G.~A. 2006{\natexlab{a}}.
\newblock A preliminary framework for description, analysis and comparison of creative systems.
\newblock \emph{Know.-Based Syst.}, 19(7): 449–458.

\bibitem[{Wiggins(2006{\natexlab{b}})}]{wiggins2006searching}
Wiggins, G.~A. 2006{\natexlab{b}}.
\newblock Searching for computational creativity.
\newblock \emph{New Generation Computing}, 24(3): 209--222.

\bibitem[{Zhang(2024)}]{zhang2024harmonyassist}
Zhang, K. 2024.
\newblock \emph{harmonyAssistant(Aosisto)}.
\newblock Master's thesis, New York University Tandon School of Engineering.

\end{thebibliography}

\end{document}